\documentclass[12pt,dvips]{article}
\usepackage{epsfig}
\usepackage{amssymb}

\title{
Fixed point action for the massless lattice 
Schwinger model\thanks{
Supported by Fonds zur F\"orderung der Wissenschaftlichen
Forschung in \"Osterreich, Project P11502-PHY.} }
\author
{\bf C.~B. Lang\footnote{e-mail: cbl@kfunigraz.ac.at}
~and T.~K. Pany \\~\\
Institut f\"ur Theoretische Physik,\\
Universit\"at Graz, A-8010 Graz, AUSTRIA}
\begin{document}
\thispagestyle{empty}
\date{31 July 1997}
\maketitle

\begin{abstract}
We determine non-perturbatively the fixed-point action for fermions in
the two-dimensional U(1) gauge (Schwinger) model. This is done by
iterating a block spin transformation in the background of non-compact
gauge field configurations sampled according to the (perfect) Gaussian
measure. The resulting action has 123 independent couplings, is
bilinear in the Grassmann fields, gauge-invariant by considered the
compact gauge transporters and localized within a $7\times 7$ lattice
centered around one of the fermions.  We then simulate the model at
various values of $\beta$ and compare with results obtained with the
Wilson fermion action. We find excellent improvement for the studied
observables (propagators and masses).
\end{abstract}

\vspace*{0.5cm}

\newpage

\section{Introduction and Motivation}

It is expected that lattice actions, which lead to continuum theories
at critical points, obey universality.  However, since computer studies
near critical points are plagued by various obstacles like finite size
effects and critical slowing down, it makes sense to work with lattice
actions that reproduce general continuum properties in some sense
faster. Technically speaking, these actions have smaller corrections in
powers  ${\cal O}(a^n)$ of the lattice spacing constant. Optimally,
such an action has no such corrections and thus no corrections to the
leading critical behaviour, i.e. no corrections to scaling. Such
actions have been called ``perfect''.

The traditional lattice actions suggested for gauge theories are simply
confined to a few lattice field variables (are ultra-local), but have
corrections ${\cal O}(a^2)$ for bosons and even larger corrections
${\cal O}(a)$ for fermions. The improved actions have to introduce more
terms. As long as the contributions are exponentially damped with
regard to their extension in real space one calls the action local.
More terms complicate the simulation and eventually one has to find a
compromise between efficiency and perfectness. Most of the improvement
programs add to traditional actions a few terms adjusting the couplings
such as to cancel lattice correction in some ${\cal O}(a^n)$. This was
done within lattice perturbation theory \cite{Sy83,ShWo85} or relying
on non-perturbative methods for determining the weight parameters for
the various terms in the action \cite{JaLiLu96}. There are also
non-perturbatively motivated suggestions for tadpole-improved actions
\cite{LeMa93}.

The other group of improvement programs is inspired by scale
transformations, in particular block spin transformations (BST).
Asymptotically free theories have their continuum limit at vanishing
gauge coupling. In such a situation one may identify a fixed point (FP)
of a BST by the solution to the classical field equations of the
combined action and BST of the spin model \cite{HaNi94}. For quadratic
actions the fixed point may be obtained quasi analytically (cf.
\cite{BeWi74} for the Gaussian model and \cite{Wi93} for free fermions
and gauge fields). Another elegant approach is direct blocking from the
continuum theory \cite{BiWi96,KeMaPa96}.

The lattice actions obtained in this way have the remarkable property
that they are {\em classically perfect}, in the sense that the
solutions of the equations of  motion are related to their continuum
counterparts.  Furthermore the FP-action is tree-level
Symanzik-improved to all orders in the lattice spacing $a$ \cite{Ni97}
and there is evidence, that cut-off effects are strongly reduced also
at the 1-loop-level \cite{DeHaHa95a,FaHaNi95,HaNi97}.

Our approach belongs to the second mentioned type of programs. For the
2D U(1) gauge theory with Wilson fermions (the lattice Schwinger model)
we determine non-perturbatively an optimal fermion action in the
background of gauge field configurations sampled according to their
(optimal) Gaussian measure.  We choose the massless Schwinger model as
our testing ground since there  we have some experience and the
possibility to compare with other results.  FP-actions for that model
were also studied by \cite{BiWi96,BiWi96a} with the method of small
fields {and recently by \cite{FaLa97} in a perturbative expansion.

We are interested in some lattice representation of the continuum 
action for the massless model \cite{Sc62b}
\begin{equation}
\frac{1}{2}\int d^2 x\,F(x)^2 - \int d^2x\,\bar{\Psi}(x)\gamma_\mu
\left(\partial_\mu+igA_\mu(x)\right)\Psi(x)\;.
\end{equation}
with one  or two flavours of fermions.  The lattice action should
respect the basic symmetries like gauge invariance, translational and
rotational invariance, parity symmetry, charge conjugation and the
hermitian invariance and should have the correct (naive)  classical
limit. Also one  has to take care for the fermion doubling problem.
Beyond these requirements the form of the lattice action is largely
arbitrary.

Here we use a real space BST with a blocking factor of 2 (Sec. 2). We
show how one can simplify the BST in the limit $\beta\to \infty$, and
with this approximation we iterate the BST to determine a FP-action.
The parameters are chosen  such that the BST yields the most local
action in the  non-interacting case.  Since we use the non-compact
formulation, the gauge field part of the FP-action may be determined
analytically \cite{BiWi96}. The fermionic part has to be determined
numerically.  The resulting FP-action in this approximation defines a
classically perfect action and for large $\beta$ we expect that it is a
good approximation for the renormalized trajectory. In this limit the
fermions decouple from the gauge field part of the BST, and  the gauge
field acts like a background field for the fermionic
sector\cite{BiBrCh96}, therefore the fermionic action stays quadratic
in the  fermionic field variables.

We suggest a parameterization for the fermionic action with terms in a
$7\times 7$ square on the lattice, using compact link variables. With
this parameterization and the Wilson fermion action as starting point
we determine the FP-action by iterating the BST.  In each step we
generate 50 coarse gauge field configurations taken from their
distribution defined by the gauge field FP-action. In the background of
these gauge fields we block the fermions from a $14\times 14$ lattice
down to the $7\times 7$ lattice and determine the iterated fermionic
couplings.

In the limit of small $A_\mu(x)$ one can compare our FP-action with the
one obtained by perturbative methods; for the so-called clover term our
results agree up to a few percent with those published in
\cite{BiBrCh96}.  We then simulate our improved action at various
values of $\beta$ for both, the 1- and the 2-flavour Schwinger model
(sec. 3). We find substantial improvement of the rotational invariance
of the mesonic 2-point function, continuum-like dispersion relations
and excellent results for the bound state masses.

\section{Discussion of the Method}

We denote the lattice action by
\begin{equation}
\beta\,S(A)-S(\bar{\Psi},\Psi,A)\;,\quad S(\bar{\Psi},\Psi,A)= 
\bar{\Psi}\,M(A)\,\Psi \;,
\end{equation}
where $S(A)$ denotes the gauge field part, $M$ the lattice Dirac
operator matrix, and $\beta=1/g^2$ is the gauge field coupling.

We block from a so-called fine square lattice with sites $x\in
Z_N\times Z_N$ to a coarse lattice organizing the fine lattice in
$2\times 2$ blocks. These blocks constitute the points $x'$ of the
coarse lattice.  We enumerate the sites $x'$ by pairs of {\em odd}
numbers such that a site $x'$ corresponds to the block $(x', x'+\hat1,
x'+\hat2, x'+\hat1+\hat2)$ on the $x$-lattice.  The Grassmann fields
are $\bar{\Psi}(x), \Psi(x)$  (respectively $\bar{\Psi}'(x'),
\Psi'(x')$ on the coarse lattice); the non-compact gauge fields
$A_\mu(x) \in \mathbb{R}$ live on the links $(x,x+\hat{\mu})$.  For the
fermions we use anti-periodic boundary conditions and for the gauge
field periodic ones.

The BST is defined as
\begin{equation}
\begin{array}{l}
\label{eq:bst}
\displaystyle
e^{ -\beta'S'(A') + S'(\bar{\Psi}', \Psi', A' )+ c } = \\
\quad\quad=
\int D_fA\,D\Psi\, D\bar{\Psi} \,e^{ -\beta\left(S(A)+T(A,A')\right) +
S(\bar{\Psi},\Psi,A)+T(\bar{\Psi},\Psi,\bar{\Psi}',\Psi') }\;,
\end{array}
\end{equation}
where $c$ is an irrelevant constant and $D_fA$ is the measure for the
gauge field including a local gauge fixing.

We fix the gauge within each block. A gauge field configuration on the 
fine lattice $A_\mu(x)$ is in the so-called {\em fine} gauge if and
only if
\begin{equation} \label{eq:finegauge}
\forall x':\quad A_1(x') = -A_1(x'+\hat 2) = - A_2(x') = A_2(x'+\hat 1)
\end{equation}
is valid (cf. Fig. \ref{FigGauge}). It can be shown by explicit
construction that for each gauge field configuration $A_\mu(x)$ there
exists a unique gauge field configuration $\hat{A}_\mu(x)$ in the fine
gauge which is related to the first one through a fine gauge
transformation. These are gauge transformations, which leave the
BST-Kernel $T(A,A')$ (for fixed $A'$, see below) invariant.

\begin{figure}[ht]
\begin{center}
\epsfig{file=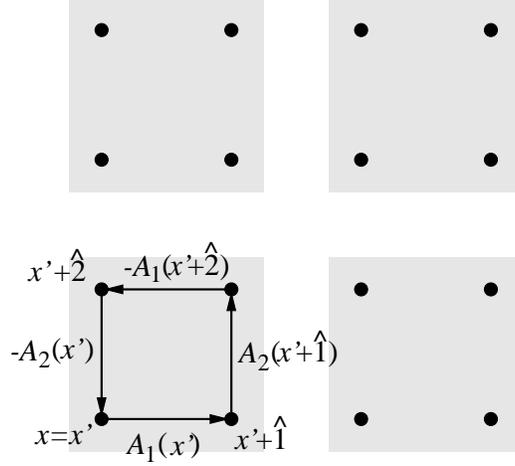,width=7cm}
\caption{\label{FigGauge}In the {\em fine} gauge the plaquette field 
strength  within each block  is distributed equally among the four link 
variables.}
\end{center}
\end{figure}

To get a well defined path integral in the BST and to apply our method
we have to get rid of these fine gauge degrees of freedom.  For that
reason we integrate not over all gauge field configurations but only
over those which are in the fine gauge.
\begin{equation}
\int D_fA\quad := \int_{
\begin{array}{c}
\displaystyle
A_\mu(x)\;\textrm{in}\cr\textrm{the fine gauge}\end{array}}DA\;.
\end{equation}

The kernel of the BST for the fermions was taken from
\cite{BiWi96,BiWi96a},
\begin{equation}
\begin{array}{l}
\label{eq:schw_bstferm}
\displaystyle
T(\bar{\Psi},\Psi,\bar{\Psi}',\Psi') = \\
\displaystyle\quad\quad\quad
\frac{1}{a_2}\, \sum_{x',i}
\left(\bar{\Psi}'_i(x') - \frac{b_2}{4}\,\sum_{x \in x'}
\bar{\Psi}_i(x) \right)
\left(\Psi'_i(x') - \frac{b_2}{4}\,\sum_{x \in x'} \Psi_i(x) \right)\;,
\end{array}
\end{equation}
where one has has to choose $b_2=\sqrt{2}$ in order to have a fixed point 
for this BST and $a_2=1/4$ for maximum locality in the situation of 
free fermions. Here $\sum_{x \in x'}$ denotes the sum over all
fine lattice sites $x$ belonging to the coarse lattice site $x'$.

For the kernel of the gauge field we define an average over the four
2-link connections between corresponding sites in adjacent blocks,
\begin{equation}
B_\mu'(x') = \frac{\beta_2}{8}\,\sum_{x \in x'}
\left(A_\mu(x)+A_\mu(x+\hat{\mu})\right)\;.
\end{equation}
A lattice differential operator of second order may be defined as
\begin{equation}
\Delta_\mu B_\nu'(x') = \delta_{\mu\nu}
\left( 2 B_\mu'(x') - B_\mu'(x'-2\hat{\mu}) - B_\mu'(x'+2\hat{\mu})
\right)\;.
\end{equation}
With these conventions we write the BST for the gauge fields as follows:
\begin{equation}
T(A,A') =  \frac{1}{2}\sum_{x',\mu}
\left( A'_\mu(x')- B_\mu'(x')\right) (\alpha_2 + \gamma_2 \Delta_\mu)^{-1}
\left( A'_\mu(x')- B_\mu'(x')\right)\;.
\end{equation}
For the wave function renormalization factor $\beta_2$, we choose the
value 2 (as in the compact case of \cite{BiWi96}) in order to make the
action on the coarse lattice gauge invariant.  This value is different
from the one which would follow due to dimensional considerations and,
in fact, for $\beta_2=2$ there exists only a quasi-FP for the BST of
the  gauge field. For this value  of $\beta_2$ for ultra-locality one
has to choose $\alpha_2 = 1/2$ and $\gamma_2 = - 1/32$.

With these values the BST respects the basic symmetries of the
Schwinger model with Wilson fermions.  The resulting action on the
coarse lattice is gauge invariant, hermitian invariant, invariant under
the charge conjugation and respects the lattice symmetry. It does
violate chiral symmetry.

For given $A'$  there exists a unique minimizing configuration of
$S(A)+T(A,A')$ which we  denote by $A_{\textrm{\scriptsize min}}(A')$. Since we use a
non-compact gauge field with action and BST quadratic in the fields,
$A_{\textrm{\scriptsize min}}(A')$ can be computed straightforwardly by solving a set
of linear equations. For $\beta \to \infty$ the saddle point
$A_{\textrm{\scriptsize min}}(A')$ dominates the path integral of the gauge field
giving
\begin{equation}
\begin{array}{l}
\label{eq:bst_1}
\displaystyle e^{ -\beta'S'(A') + S'(\bar{\Psi}', \Psi', A' )+c } =
e^{ -\beta\left(S(A_{\textrm{\scriptsize min}}) + T(A_{\textrm{\scriptsize min}},A')\right) + c_1 }
\times \\
\displaystyle \times\left(
\int D\Psi D\bar{\Psi}\, \exp\left[
S(\bar{\Psi},\Psi,A_{\textrm{\scriptsize min}})+T(\bar{\Psi},\Psi,\bar{\Psi}',\Psi')
\right] 
+{\cal O}\left(\frac{1}{\beta}\right)\right)\;,
\end{array}
\end{equation}
where $c_1$ is some constant which may be absorbed into $c$.  For
Grassmann variables the ``gaussian''  integral results in the
exponential of an element of the Grassmann algebra in $\Psi'$ and
$\bar{\Psi}'$.  This algebra element is a sum of bilinear terms in the
fermionic variables plus a constant. It may therefore be identified
with the blocked fermionic action by determining the coefficients of
the corresponding fermion matrix.  Formally this is equivalent to the
saddle point minimization for the bosonic action.  Due to this analogy
to bosons we will denote this process of integration over the Grassmann
fields and subsequent identifications of the coefficient of the
fermionic action by $\min_{\{\bar{\Psi},\Psi\}}$.

Having solved the two path integrals (the bosonic and the formal
fermionic one) we separate the resulting action of (i.e. the logarithm
of (\ref{eq:bst_1})) into  two parts, one belonging to the scalar
subspace of the Grassmann-algebra, the other one quadratic in the
fermionic fields.
\begin{eqnarray}
\beta'S'(A') &=& \beta\left(S(A_{\textrm{\scriptsize min}}) 
+ T(A_{\textrm{\scriptsize min}},A')\right) + c_2(A_{\textrm{\scriptsize min}}) +
{\cal O}\left(\frac{1}{\beta}\right)\;, \nonumber\\
S'(\bar{\Psi}', \Psi', A' ) &=&  \min_{\{\bar{\Psi},\Psi\}}
\left( S(\bar{\Psi}, \Psi, A_{\textrm{\scriptsize min}})	+
T(\bar{\Psi},\Psi,\bar{\Psi}',\Psi') \right)+
{\cal O}\left(\frac{1}{\beta}\right)\;.
\end{eqnarray}
The constant $c_2(A_{\textrm{\scriptsize min}})$ is just the logarithm of the
fermionic determinant resulting from the Grassmann integral.
Compared to the leading term proportional to $\beta$
we may neglect this contribution and find that
the fermions decouple completely from the gauge field BST. 
The defining equations for the FP-action finally have the form
\begin{eqnarray} \label{eq:bst_gauge}
\beta'S_*(A') &= \beta\, \min_{A\;\textrm{in the fine gauge}}
\left(S_*(A) + T(A,A')\right)\;, \\
S_*(\bar{\Psi}', \Psi', A' ) &=  \min_{\{\bar{\Psi},\Psi\}}
\left( S_*(\bar{\Psi}, \Psi, A_{\textrm{\scriptsize min}}) +
 T(\bar{\Psi},\Psi,\bar{\Psi}',\Psi') \right)\;.
\label{eq:bst_ferm}
\end{eqnarray}
These equations replace now the BST at $\beta \to \infty$. Since we
want to use the action also at moderate $\beta$ values, we have to
calculate $S'$ for strongly fluctuating configurations $A'$, too. This
however is naturally possible, since we never demanded that $A'$ ought
to be small. As was already mentioned in \cite{HaNi94} this method has
nothing to do with perturbation theory. It can be shown that the form of
the action is independent of the number of fermion species.

Here we should mention that both, the FP-action for the gauge field and
the FP-action for the fermions do have scale invariant solutions.  For
the gauge-field this was proven in \cite{HaNi94}; for a non-compact
gauge theory with periodic boundary conditions this is not very
interesting, since there are no instanton solutions. For the fermionic
FP-action the  theorem takes the following form. Suppose that
$\bar{\Psi}'$ and $\Psi'$ are  solutions of the classical equations of
motion $\partial S(\bar{\Psi}',\bar{\Psi},A')/\partial \bar{\Psi}' = 0$
and $\partial S(\bar{\Psi}',\bar{\Psi},A')/\partial \Psi' = 0$, i.e.
are part of the null-space of the fermionic matrix, then also the
minimizing configurations of (\ref{eq:bst_ferm}) $\bar{\Psi},\Psi$ on
the fine lattice are solutions of the equations of motion corresponding
to $S(\bar{\Psi},\Psi,A_{\textrm{\scriptsize min}}(A'))$.

In the following two sections we discuss how to solve
(\ref{eq:bst_gauge}) by analytical methods and (\ref{eq:bst_ferm}) by
numerical methods.

\subsection{The FP-action for the gauge field}

For details we refer to \cite{BiWi96a,BiWi96}.  For our choice of the
parameters of the BST the ultra-local standard (non-compact) plaquette
action is a fixed point, up  to the wave  function renormalization.  In
$d=2$ this action is
\begin{eqnarray}\label{eq:stdplaq}
S_P(A)&=&\frac{1}{2}\,\sum_x F(x)^2\nonumber \\
&=& \frac{1}{2}\,
\sum_x\left( A_2(x+\hat 1) - A_2(x) -
A_1(x+\hat 2) + A_1(x)\right)^2\;.
\end{eqnarray}
As it was shown in \cite{BiWi96} on an infinite lattice the equation
\begin{equation} \label{eq:bst_res_gauge}
\frac{1}{4}S_P(A') =  \min_{\{A\}}\left(S_P(A)+T(A,A')\right)
\end{equation}
is valid.  It is remarkable, that (\ref{eq:bst_res_gauge}) is also
exactly fulfilled on a finite lattice as long as $S_P$ fits on the
coarse lattice.  For this reason we forget about the  finite size
effects even in the  case of the FP-action for the fermions (cf. also
\cite{DeHaHa95b}).

Although $S_P$ is strictly speaking not a  fixed  point under the BST
(due to the necessary but trivial rescaling of $\beta$) it is still a
perfect action. In the Schwinger model it is classically perfect and
for the free gauge field it is quantum perfect, since it describes the
same physics as on an infinite  fine lattice. To determine the
FP-action for the fermions it is  necessary to iterate the BST and
therefore it is necessary to renormalize $\beta' \to 4 \beta$ after
each  step to avoid the unwanted convergence of $\beta\to 0$.

\subsection{The FP-action for the fermions}\label{fpalgorithm}

The fixed point  action for the fermions can be calculated  in a
perturbation expansion as was shown in \cite{BiWi96,BiWi96a} or more
explicitly in \cite{BiBrCh96}. But in \cite{BiFoWi95a} the authors
demonstrate for the case of the Gross-Neveu model at large $N$ that the
non-perturbative approach of \cite{HaNi94} is really necessary  to
calculate the perfect action.

Eq. (\ref{eq:bst_ferm}) defines the action on the coarse lattice for
any configuration $A'$, and -- as we discuss later -- one can advise
numerical procedures  which calculate the action to high precision.
However, since we want to iterate the BST and since we want to use this
action in numerical simulations we  have to parameterize
$S'(\bar{\Psi}',\Psi',A')$ with a finite number of coupling constants.
\begin{equation} \label{eq:fermfit}
S_F(\bar{\Psi},\Psi,A) = \bar{\Psi} M_F(A) \Psi =
\sum_{i=0}^3\sum_{x\, , f}\,  \rho_i(f)\, 
\bar{\Psi}(x) \sigma_i U(x,f) \Psi(x+\delta f)\;.
\end{equation}
Here $M_F(A)$ is the parameterized fermion matrix. By $f$ we denote a
closed loop through $x$ or a path from the lattice site $x$ to
$x+\delta f$ ($\delta f$ denotes the distance vector on the lattice
corresponding to the path $f$).  $U(x,f)$ is the parallel transporter
(i.e. the element of the compact group) along this path. The
$\sigma_i$-matrices denote the Pauli matrices for $i= 1,2,3$ and the
$2\times 2$ unit matrix for $i=0$.

We have to ensure that the BST stays  on  the critical surface and
eventually converges to the nontrivial fixed point.  Thus we impose the
condition
\begin{equation}
\sum_f \rho_0(f) = 0\;,
\end{equation}
which guarantees, that  $S_F$
reproduces the action of the massless continuum Schwinger model
in the naive continuum limit.
The normalization of $\rho$ is fixed by demanding 
\begin{equation}
\sum_f \rho_1(f) (\delta f)_1 = 1\;.
\end{equation}
Here $(\delta f)_1$ denotes the component of the path vector  in the
1-direction.  We want to emphasize that (\ref{eq:fermfit}) is not the
most general parameterization. E.g., since the gauge fields enter as
parallel transporter in their compact version a clover-term
$\bar{\Psi}(x) F_{\mu\nu}(x) \Psi(x)$, which contains the non-compact
field strength, can be represented only in an approximate way for
values $-\pi < F(x) < \pi$.  For sufficiently large values of  $\beta$
this should be no problem.

We require the  invariance  of  the action under  certain  symmetries;
thus we can impose some conditions for the coupling constants
$\rho_i(f)$. For example hermitian invariance and invariance  under
charge conjugation implies that $\rho_{0,1,2}(f)$ has to be real and
$\rho_3(f)$ has to be purely imaginary. These and further symmetries
due to the lattice geometry drastically reduce the number of
independent coupling constants.  We have considered terms that connect
the central site $x$ with any other site $x+\delta f$ in a $7\times 7$
lattice. Concerning the length of the connecting paths $f$ we first
considered paths which may exceed the shortest connection between $x$
and $x+\delta f$ by up to 8 links and some extra paths containing
higher powers of plaquette variables. However, in the iteration
procedure it turned that one may omit many of these.  Altogether,
respecting the mentioned symmetries, we finally considered 33 different
geometric shapes corresponding to 123 independent coupling constants
\cite{WWW}.

For the determination of  the fixed point action we proceed as follows.
The starting point for the fermionic action is the Wilson action for
the massless Schwinger model ($\kappa  = 1/4$). 
\begin{enumerate}
\item
We generated 50 gauge field configurations $A'$ on the coarse $7\times
7$ lattice according to their probability distribution $e^{-\beta'
S_P(A')}$ (with periodic boundary conditions).  For definiteness this
has to be done in a fixed gauge.  Technically, we do this by randomly
sampling the non-zero diagonal elements of the momentum space
representation of the action \cite{ChWe79}.  For each of these
configurations we then calculate the corresponding minimizing
configuration $A_{\textrm{\scriptsize min}}(A')$ as  described above.
\item
With the help of $A_{\textrm{\scriptsize min}}(A')$ we generate the fermion matrix on
the fine lattice and performed the BST (Grassmann integral) for the
fermions giving  the $(2\cdot7^2)\times(2\cdot7^2)$ fermion matrix
$M_{BST}(A_{\textrm{\scriptsize min}})$ on the coarse lattice. This is done for all
50 gauge field configurations.
\item
The resulting fermion matrices are then compared with the
fermion matrices $M_F(A')$ for the coarse lattice generated for the
corresponding $A'$ configurations. A new set of parameters according
(\ref{eq:fermfit}) is now determined by minimizing
\begin{equation}
\sum_{A'} \| M_{BST}\left(A_{\textrm{\scriptsize min}}(A')\right)-M_F(A')\|^2\;,
\end{equation}
where the matrix norm is defined
\begin{equation}
\|M\|^2\equiv\sum_{i,j}\,|M_{ij}|^2\:.
\end{equation}
\end{enumerate}
All these steps are iterated until the coupling constants remain stable
within  small statistical fluctuations.

We worked at $\beta' = 20$ which corresponds to a typical value for the
field strength $F(x)\approx 0.18$.  After ${\cal O}(10)$ iterations the
coupling constants stabilized and subsequent iterations stayed within
the statistical errors.  The largest matrix element is $\sim 1.4$; the
average deviation between individual matrix elements given by  the BST
and by the parameterization was $0.00024$ in the last iteration step.
Comparing different sets of $A'$ the values of the coupling constants
we obtained in this way  differed in the mean by $0.0017$, which
provides an estimate of the possible systematic uncertainty due to the
finite number of configurations. We have no control on possible
redundancies in the parameterized action. Thus it may well be, that
part of the observed (small) fluctuations in the couplings is due to
cancellations of certain terms in the fermionic action. The final
action therefore may have been determined to an even higher accuracy.

\begin{figure}[htb]
\begin{center}
\epsfig{file=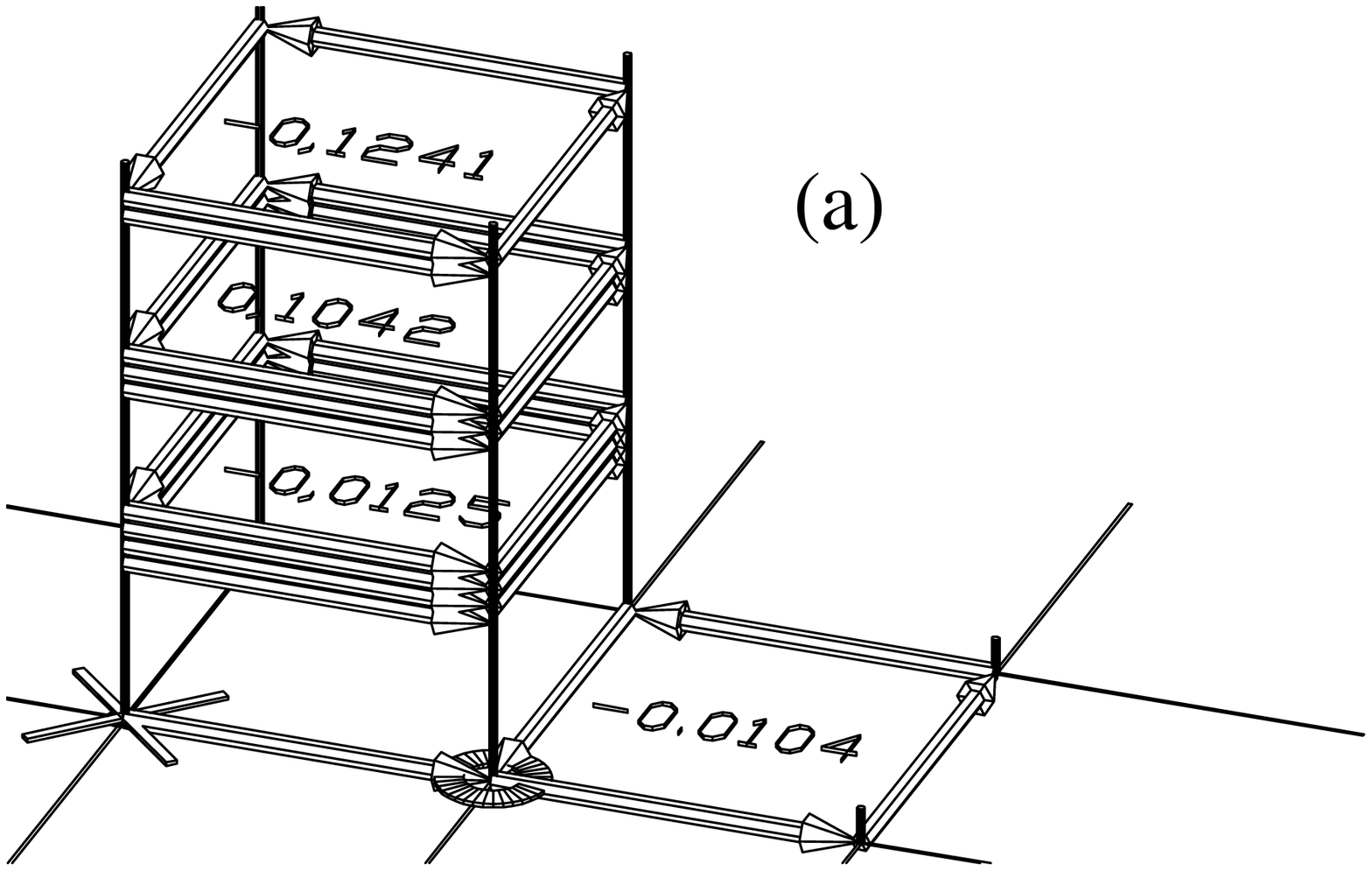,width=7cm}
\hfill
\epsfig{file=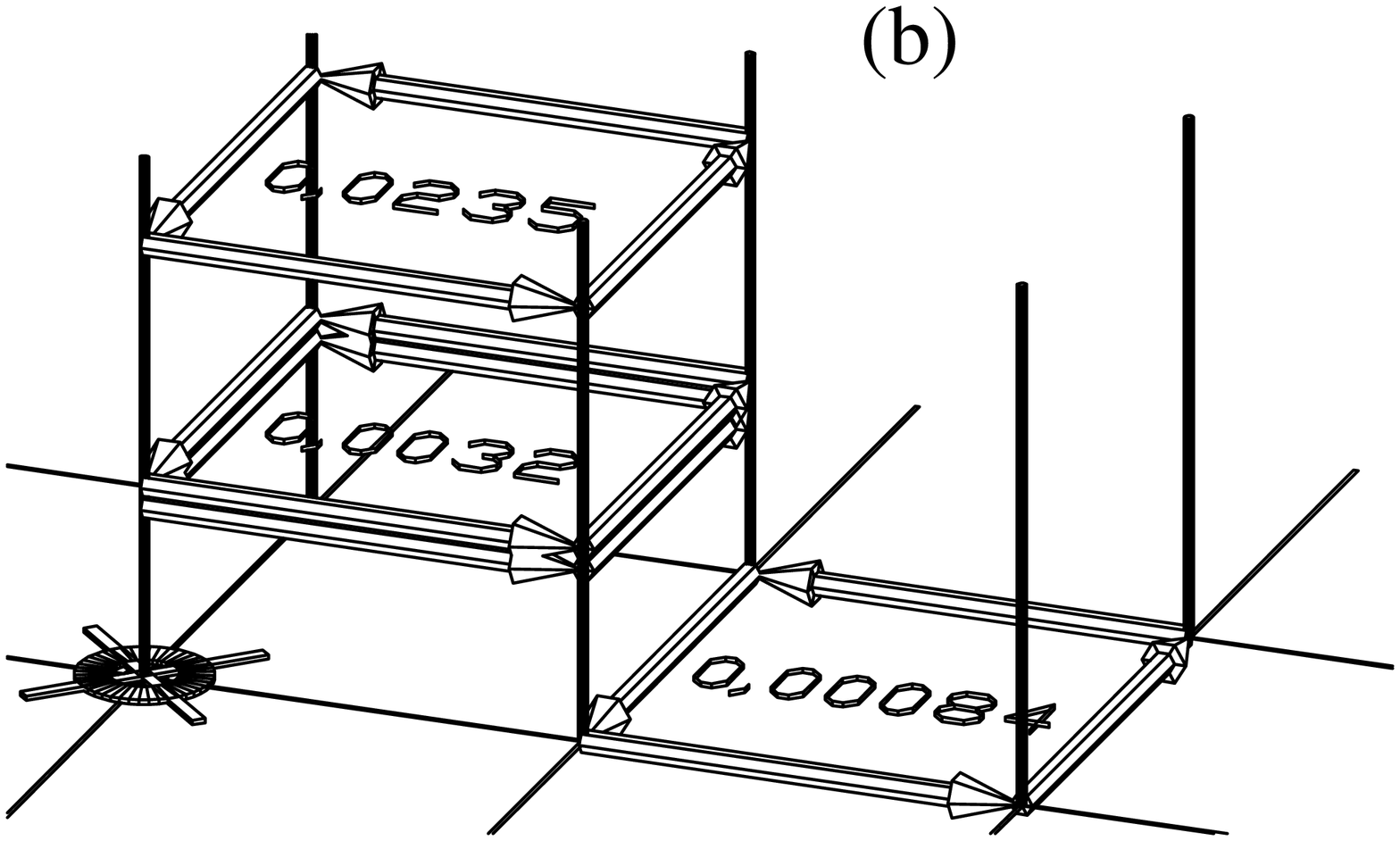,width=6cm}
\end{center}
\caption{
\label{fig:act_res} (a) Symbolic representation of some of the
$\sigma_3$ couplings of the  perfect action.  Crosses and circles mark
the position of the $\bar{\Psi}$ and $\Psi$.  Each block of arrows
corresponds to a term $\rho_3(f)\,\bar{\Psi}(x)\,\sigma_3 \,U(x,f)\,
\Psi(x+\delta f)$, where $f$ is symbolized by  arrows and the coupling
$\rho_3(f)$ is written at the height of the arrows.  (b) Terms building
the perfect clover term.}
\end{figure}

In figure \ref{fig:act_res} we show a subset of the couplings of the
FP-action, which is proportional  to $\bar{\Psi} \sigma_3 \Psi$. In the
limit of small  $A_\mu(x)$ one can expand $U(x,f)$ and the expressions
linear in $A_\mu(x)$ can be  compared with Fig. 14 of \cite{BiBrCh96}.
If we compare the two most important operators of this perfect  clover
term, our result agrees within a few percent with \cite{BiBrCh96}. The
smaller coupling constants of the clover term have already rather large
statistical errors.  The locality of our FP-action is established by
comparing our couplings with the situation of vanishing gauge fields.
We find that our values are similar or smaller than those obtained for
the free fermion perfect action \cite{BiWi96}.  The complete set of
couplings may be retrieved from \cite{WWW}.

\section{Simulation of the FP-action}\label{se:results}

For notational simplicity we call now the obtained parameterization of
the FP-action just {\em the} FP-action.  For $\beta\to\infty$ the
FP-action is on the RT by construction. For finite values of $\beta$ it
may be no longer perfect. This depends on the system considered; the
FP-action of the O(3)-model turned out to stay close to perfectness
\cite{HaNi94}.  In order to check the amount of improvement, we have to
rely on direct simulations with the FP-action. Indeed, as will be shown
below, we found significant improvement for all observables studied.

One has to determine the path integral
\begin{equation}\label{eq:simul_pfad}
\frac{1}{Z}\int DA\,\int D\Psi D\bar{\Psi}\, {\cal{O}}(\bar{\Psi},\Psi,A)
e^{-\beta S_P(A) + S(\bar{\Psi},\Psi,A) }\;.
\end{equation}
A full-scale (hybrid-) Monte-Carlo simulation for the FP-action is a highly 
non-trivial task. We decided to work on lattices of moderate size
up to $16\times 16$. In this case one may perform the Grassmann
integrals explicitly, i.e. by computing the corresponding determinant
and inverse fermion matrix.

For a simulation we generated $5000$ gauge field configurations with
periodic boundary conditions according to their gaussian probability
distribution $e^{-\beta S_P(A)}$ with gauge fixing as has been
discussed above in Sec. \ref{fpalgorithm} in the context of the BST.
For these configurations we determined the fermionic path integral as
discussed with standard routines of linear algebra.

The side-benefit of this approach is that we obtain results for both,
the determinant corresponding to the 1-flavour model and its square,
corresponding to the 2-flavour situation. All observables are obtained
in that way.  In order to estimate the statistical errors we repeated
the whole procedure several times (several runs of 5000 gauge
configurations each).  We compared the results obtained with the
FP-action with results obtained in in the same way for the
Wilson-action for fermions (but still using the non-compact form of the
gauge field action). For the Wilson action we use $\kappa=0.25$
throughout. This is below the critical value $\kappa_c(\beta)$ which,
however, approaches 0.25 asymptotically. We therefore do not expect
perfect agreement of the obtained physical mass values for thw Wilson
action. However, for a qualitative comparison of improvement properties
this choice should be sufficient.

\begin{figure}[htbp]
\begin{center}
\epsfig{file=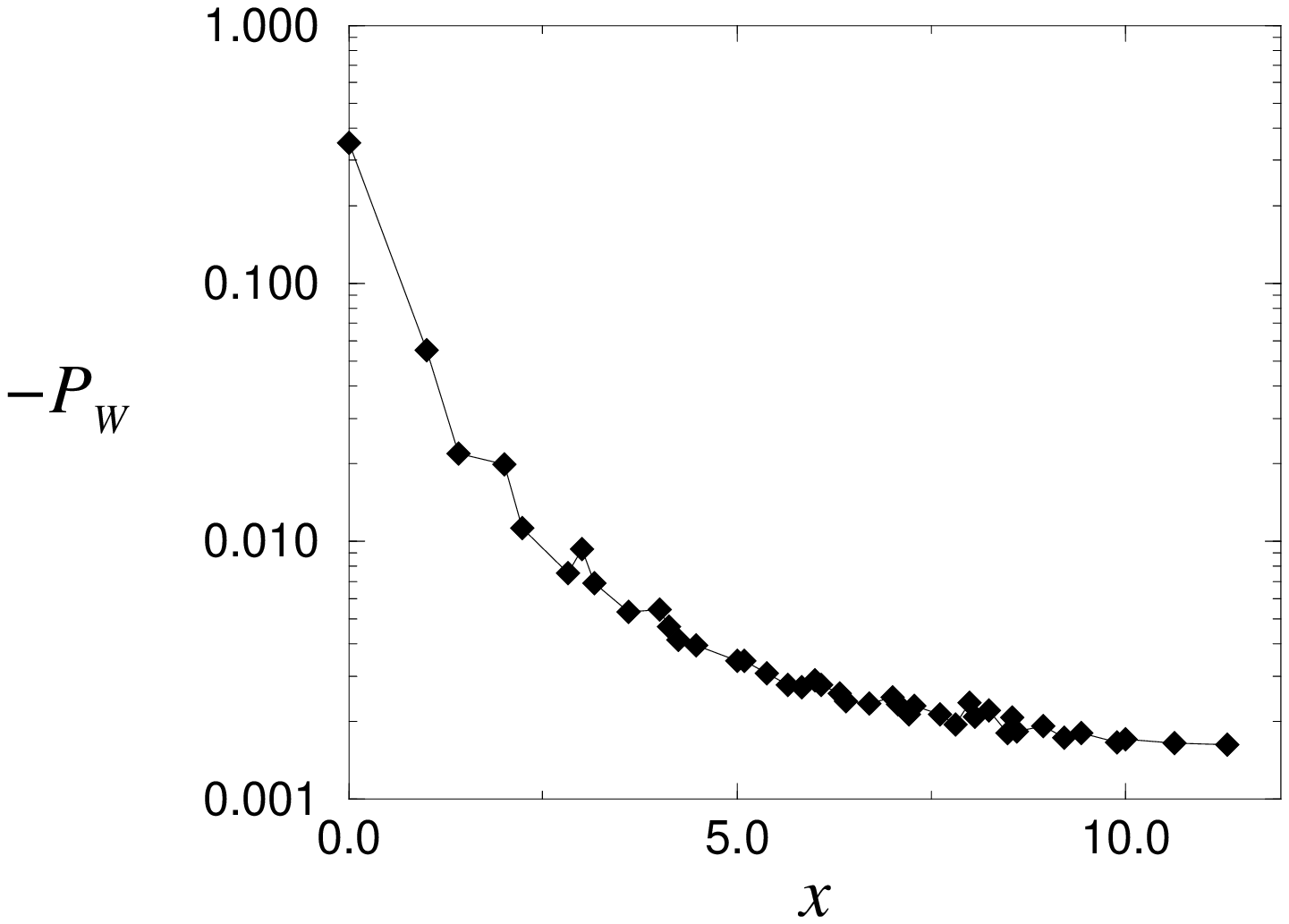,width=6.5cm}
\hfill
\epsfig{file=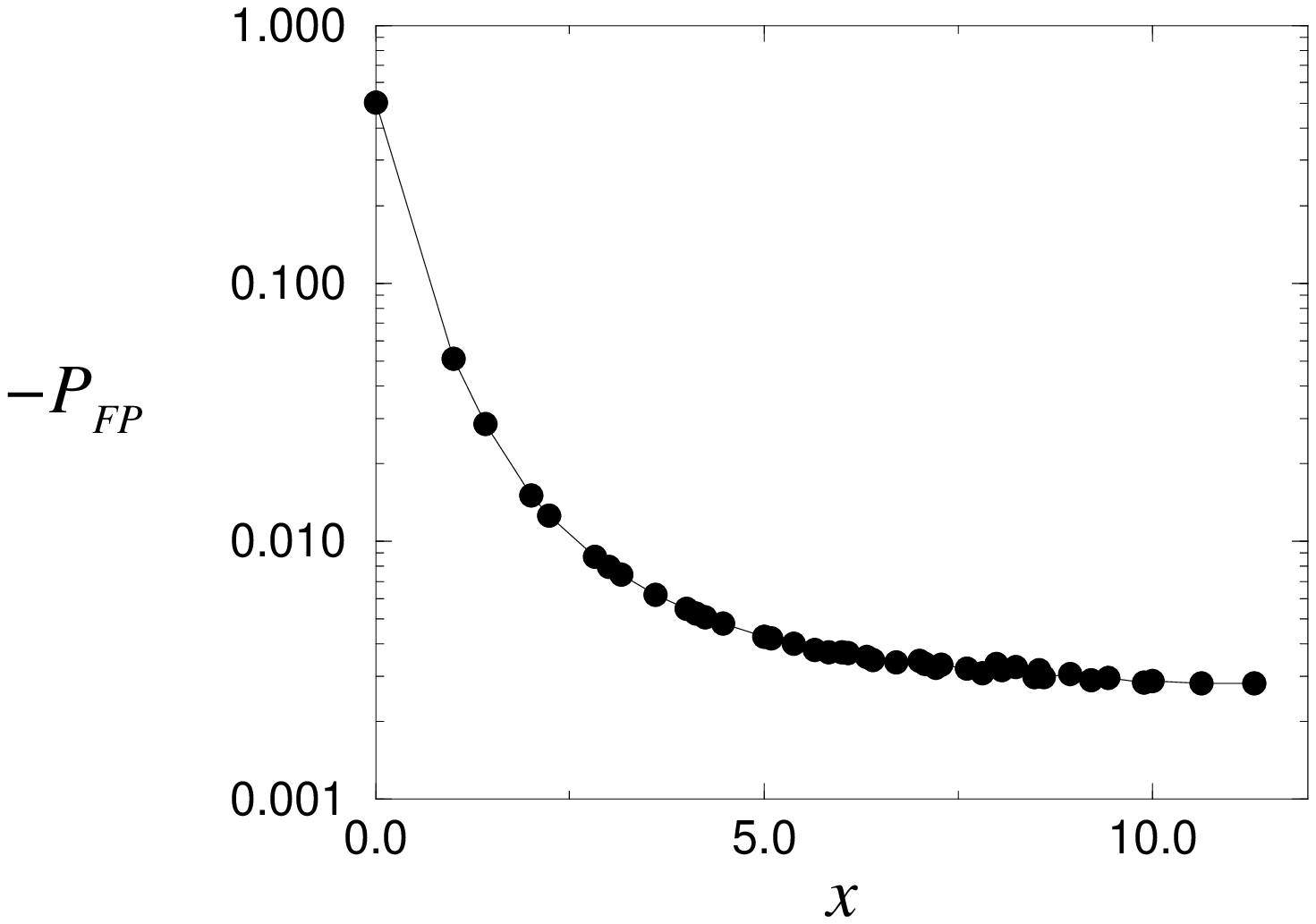,width=6.5cm}
\caption{Improvement of rotational invariancefor the correlation
function (\ref{eq:corfun}) in the 1-flavour model for (left-hand plot)
Wilson and (right-hand plot) the FP-action on a $16\times 16$ lattice
at $\beta=6$.  \label{fig:rot_fp}}
\end{center}
\end{figure}

Fig. \ref{fig:rot_fp} exhibits for the 1-flavour model the correlation 
function
\begin{equation}\label{eq:corfun}
P(x)=\langle\bar{\Psi}(0)\,\sigma_3\,\Psi(0)\,
\bar{\Psi}(x)\,\sigma_3\,\Psi(x)\rangle
\end{equation}
measured  for all 2-point separations. The numbers have been determined
at $\beta=6$. One finds much better rotational invariance for the
FP-action than for the Wilson-action.

\begin{figure}[tbp]
\begin{center}
\epsfig{file=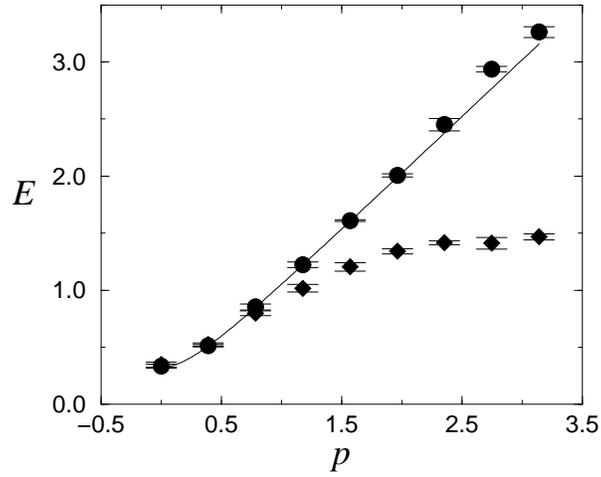,width=8cm}
\caption{Dispersion relation of the massive boson in the 1-flavour
Schwinger model with Wilson fermions (diamonds) and  with the FP-action
(full circles) calculated on a $16\times 16$ lattice at $\beta=6$,
compared with the continuum dispersion relation $E(p)=\sqrt{m^2+p^2}$.
\label{fig:disp_rel}}
\end{center}
\end{figure}

\begin{figure}[tbp]
\begin{center}
\epsfig{file=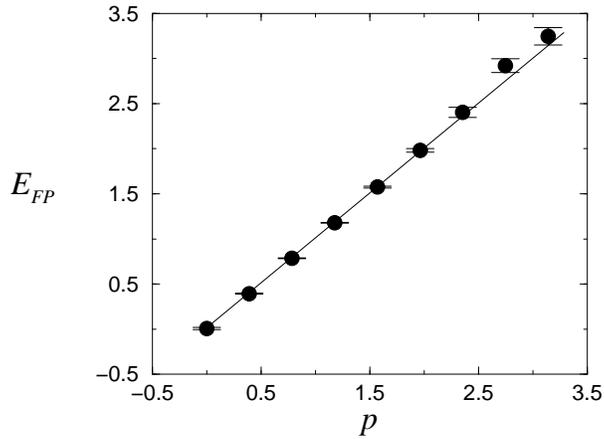,width=8cm}
\caption{Dispersion relation of the massless ($\pi$) boson in the 
2-flavour Schwinger model with the FP-action calculated on 
a $16\times 16$ lattice at $\beta=6$, 
compared with the continuum dispersion relation
$E(p)=p$.\label{fig:pi_disp_fp}}
\end{center}
\end{figure}

In the 2-flavour Schwinger model one expects one massive mode (which we
call $\eta$ by analogy) and a massless flavour-triplet (called $\pi$).
The corresponding momentum-projected operators are
\begin{eqnarray}
\eta(p,t) &=& \sum_{x_1} \,e^{ipx_1} 
\left(\bar u(x_1,t)\,\sigma_1\,u(x_1,t)
+\bar d(x_1,t)\,\sigma_1 \,d(x_1,t)\right)\;,\\
\pi(p,t) & = &\sum_{x_1} \,e^{ipx_1} 
\left(\bar u(x_1,t)\,\sigma_1\,u(x_1,t)
-\bar d(x_1,t)\,\sigma_1 \,d(x_1,t)\right)\;.
\end{eqnarray}
Their correlation functions define by their exponential decay the
corresponding energy functions $E(p)$. In Fig.\ref{fig:disp_rel} we
present results for the $\eta$ dispersion relation for the
Wilson-action and the FP-action and compare these with the continuum
dispersion relation. Again we find significant improvement.

\begin{figure}[tb]
\begin{center}
\epsfig{file=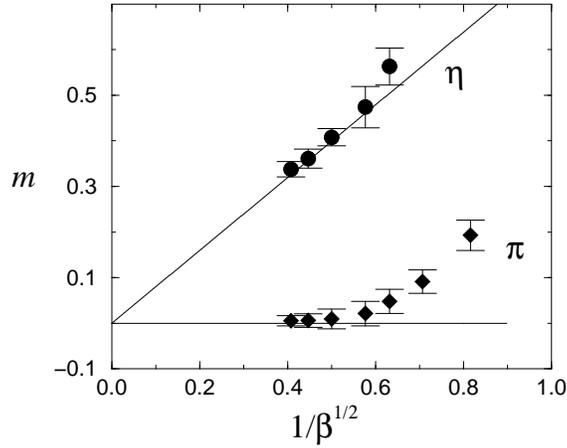,width=7.5cm}
\caption{Masses for the massive (full circles) and the massless (diamonds) modes
in the 2-flavour Schwinger model vs. $1/\sqrt{\beta}$; the values have
been determined with the FP-action on an $16\times 16$ lattice. The line
denotes the theoretical expectations for scaling.
\label{Fig_Scaling}}
\end{center}
\end{figure}

The dispersion relation for the $\pi$-state is plotted in Fig.
\ref{fig:pi_disp_fp} only for the FP-action. For the Wilson action at
$\kappa=0.25$ and $\beta=6$ this state has a non-vanishing mass and is
therefore not suitable for a comparison.

It is  remarkable that the $\pi$-mass obtained by the means of the
FP-action is even numerically very close to zero (e.g. 0.005 at
$\beta=6$), whereas with Wilson-fermions one has to search for a
critical point in $\kappa$.  In fact for $\beta\to\infty$ this
constitutes a test for the perfectness of the FP-action, since for the
optimal action of course one has to recover the massless continuum
Schwinger model properties.  Also the masses obtained for the massive
mode are within the statistical errors in agreement with the
theoretical expectations, e.g.  $m_\eta=0.33(1)$ compared to 0.3257
from continuum theory (at $\beta=1/g^2=6$).  Since the mass from the
lattice propagator has been determined by the usual $\cosh$-fit on
$16\times 16$ lattices the value has to be handled with some caution.

Fig. \ref{Fig_Scaling} shows the obtained masses for massive and
massless modes at various values of $\beta$ (determined with the
FP-action on $16\times 16$-lattices). We find clear signals of
non-vanishing $\pi$-masses at sufficiently small $\beta$, indicating
deviation of our FP-action from the renormalized trajectory. However,
the overall scaling behaviour predicted (for the 2-flavour model) from
theory,
\begin{equation}
a(\beta)\,m_\eta = \sqrt{\frac{2}{\pi\,\beta}}\;
\end{equation}
is nicely recovered for moderately large values of $\beta>3$.

\section{Conclusion}

For the 2D Schwinger model we determined the optimal FP-action for
interacting fermions in a non-perturbative background of gauge fields.
The parameterized action contains 123 independent couplings localized
in a $7\times 7$ lattice centered around one of the fermionic fields.

With this action we have simulated the model on a $16\times 16$ lattice
for various values of the gauge coupling and for one and two species of
fermions. Our results show excellent improvement of the important
propagator observables, although we did not improve those operators.
\\
~\\
{\bf Acknowledgment:} We wish to thank W. Bietenholz, F. Farchioni, I.
Hip, U.-J. Wiese and E. Seiler for discussions.

\end{document}